\documentclass[12pt]{article}
\def\bfi{\bf}
\def\setmarsing-ok{
\textwidth 6.5in
\textheight 8.5in
\oddsidemargin 0in
\topmargin -0.5in}

\setmarsing-ok

\begin{document}


\title{
Renormalized HVBK dynamics for Superfluid Helium
Turbulence }

\author{
Darryl D. Holm
\\Theoretical Division and Center for Nonlinear Studies
\\Los Alamos National
Laboratory, MS B284
\\ Los Alamos, NM 87545\\
{\footnotesize dholm@lanl.gov}\\
}
\date{Isaac Newton Institute Workshop Proceedings\\
Quantum Vortex Dynamics\\
\vspace{2mm}
Cambridge UK, August 2000}

\maketitle
\normalsize
\vspace{-0.1in}

\begin{abstract}

We review the Hall-Vinen-Bekarevich-Khalatnikov (HVBK) equations for
superfluid Helium turbulence and discuss their implications for recent
measurements of superfluid turbulence decay.

A new Hamiltonian formulation of these equations renormalizes the vortex
line velocity to incorporate finite temperature  effects. These effects
also renormalize the coupling constant in the mutual friction force
between the superfluid and normal fluid components by a factor of
$\rho_s/\rho$ (the superfluid mass fraction) but they leave the vortex
line tension unaffected.  Thus, the original HVBK form is recovered at
zero temperature and its mutual friction coefficients are renormalized at
nonzero temperature. The HVBK equations keep their form and no new
parameters are added. However, a temperature dependent trade-off does
arise between the mutual friction coupling and the vortex line tension.
\smallskip

The renormalized HVBK equations obtained via this new Hamiltonian approach
imply a dynamical equation for the space-integrated vortex tangle
length, which is the quantity measured by second sound attenuation
experiments in superfluid turbulence. A Taylor-Proudman theorem also
emerges for the superfluid vortices that shows the steady vortex line
velocity becomes columnar under rapid rotation.\\


\end{abstract}


\vskip 4mm

\section{HVBK equations}

Recent experiments establish the Hall-Vinen-Bekarevich-Khalatnikov (HVBK)
equations as a leading model for describing superfluid Helium
turbulence. See Nemirovskii and Fiszdon [1995] and {Donnelly [1999]} for
authoritative reviews. See Henderson and Barenghi [2000] for a recent
fluid mechanics study of steady cylindrical Couette flow using computer
simulations of the incompressible HVBK equations. 

In the Galilean frame of the normal fluid with velocity
$\mathbf{v}_n$, the HVBK equations may be expressed as follows, upon
ignoring thermal diffusivity and viscosity, 
\begin{eqnarray}
&&\partial_t\rho
=
-\,{\rm div}\mathbf{J}
\,,
\qquad
\partial_t J_i
=
-\,
\partial_k \, T_i^{\,k}
\,,\qquad
\partial_t S
=
-\,{\rm div}(S\mathbf{v}_n)
+ R/T
\,,
\nonumber\\
&&
\rho_s\partial_t \mathbf{v}_s
+
\rho_s(\mathbf{v}_s\cdot\nabla)\mathbf{v}_s
=
-\,
\rho_s\nabla\big(\,\mu 
- \frac{1}{2}|\mathbf{v}_s - \mathbf{v}_n|^2\big)
+
\rho_s\mathbf{f}
\,,\label{SF-vel-dyn}
\end{eqnarray}
and summing over pair of upper and lower repeated indices.
One may consult, e.g., Bekarevich and Khalatnikov (BK) [1961] and
{Donnelly [1999]} to compare these equations with the form they take in
the reference frame of the superfluid.%
\footnote{In making this comparison it is useful to recall the Galilean
transformation of the chemical potential,
$\mu^{\,\prime} 
=
\mu 
- \frac{1}{2}|\mathbf{v}_s - \mathbf{v}_n|^2
\,,
$
where $\mu^{\,\prime}$ is evaluated in the superfluid frame and $\mu$
in the normal-fluid frame. See Putterman [1974] for a clear discussion of
the role of Galilean transformations in superfluid hydrodynamics.}

\noindent
{\bf Notation.} Here $\rho$ and $\rho_s$ denote the total and superfluid
mass densities, respectively. The entropy density of the
normal fluid is $S$, its temperature is denoted $T$, and
$\rho_n=\rho-\rho_s$ denotes its mass density. The superfluid velocity is
denoted $\mathbf{v}_s$ and
\[
\mathbf{J}=\rho_s\mathbf{v}_s + \rho_n\mathbf{v}_n
\]
is the total momentum density. In the entropy equation $R$ is the rate
that heat is produced by the phenomenological friction and reactive forces
in $\mathbf{f}$, which must be specified to close the theory. We also
denote
\begin{eqnarray}
\hbox{Stress tensor:}&&
T_i^{\,k}
=
\,\rho_s\, v_{s\,i}\,v_s^k +\rho_n\, v_{n\,i}\,v_n^k 
+
(P
+
{\lambda}\!\cdot\!{\omega})\,\delta_i^k 
-
\,\lambda_i\omega^k 
\,,\nonumber\\
\hbox{Euler's pressure law:}&&
P 
=
-\,\varepsilon_0 + TS + \rho\mu
\,,\nonumber\\
\hbox{Superfluid First Law:}&&
d\varepsilon_0 
= 
\mu\,d\rho + TdS 
+ 
(\mathbf{J}-\rho\mathbf{v}_n)\!\cdot\!d(\mathbf{v}_s-\mathbf{v}_n)
+ 
{\lambda}\!\cdot\!d{\omega}
\,.\label{1stLaw-in-normal-frame}
\end{eqnarray}
In the superfluid First Law,  ${\omega}={\rm
curl\,\mathbf{v}_s}$ is the superfluid vorticity with magnitude
$|\omega|={\hat{\omega}}\cdot{\omega}$, where
${\hat{\omega}}={\omega}/|\omega|$ is its unit vector.
BK [1961] takes the energy density $\varepsilon_0$ to depend on the
magnitude of the superfluid vorticity, $|\omega|$, as 
\begin{equation}
\varepsilon_0 = \frac{\rho_s\kappa\,|\omega|}{4\pi}{\rm ln}\,\frac{R}{a}
\,.
\nonumber
\end{equation}
This is the energy per unit length of a superfluid vortex line, 
$\rho_s(\kappa^2/4\pi){\rm ln}(R/a)$, with quantum of circulation
$\kappa=h/m\simeq10^{-3}$(cm$^2$/sec) and ratio $R/a$ of mean distance
between vortices $R$ to effective vortex radius $a$, times the vortex
length per unit volume, $|\omega|/\kappa$. Hence, we find
\begin{equation}
{\lambda} =
\frac{\partial\varepsilon_0}{\partial|\omega|} 
\frac{\partial|\omega|}{\partial{\omega}} = 
|\lambda|\,{\hat{\omega}}
\quad\hbox{with}\quad
|\lambda|/\rho_s \approx \frac{\kappa}{4\pi}{\rm ln}\,\frac{R}{a}
 = \lambda_0
\,,\nonumber
\end{equation}
where  one ignores the derivative of $R\simeq \sqrt{\kappa/|\omega|}$ 
inside the logarithm. The appearance of ${\lambda}$ in the stress tensor
$T_i^{\,k}$ shifts the pressure $P$, and div$\,T$  introduces an
additional force $-\,{\omega}\cdot\nabla{\lambda}\equiv-\rho_s\mathbf{T}$
into the motion equation. The quantity $\mathbf{T}$ is called the ``vortex
line tension.''

BK [1961] assigned the following form to the phenomenological coupling
force $\mathbf{f}$ appearing in the superfluid velocity equation in 
(\ref{SF-vel-dyn}),
\begin{eqnarray}
\mathbf{f} 
&=&
(\mathbf{v}_L-\mathbf{v}_s)\times{\omega}
\,,\qquad
\mathbf{v}_L 
=
\mathbf{v}_n
-\,
\rho_s\,( \alpha\,\mathbf{s}^0
+ 
\beta\,{\hat{\omega}}\times\mathbf{s}^0 )
\,,\label{Phenom-force-HVBK}\\
\hbox{where}\quad
\mathbf{s}^0 
&=&
\mathbf{v}_\ell^0 - \mathbf{v}_n
\,,\hspace{2cm}
\mathbf{v}_\ell^0
=
\mathbf{v}_s 
+ 
\rho_s^{-1}{\rm curl}\,{\lambda}
\,,\label{vort-vel-zero}
\end{eqnarray}
with ``vortex velocity'' $\mathbf{v}_L$ and ``slip velocity''
$\mathbf{s}^0$ introduced as auxiliary quantities. The HVBK equations
written as (\ref{SF-vel-dyn}) in the normal-fluid frame conserve the
energy,
\begin{equation}\label{HVBK-erg}
E 
=
\int
\Big[
\frac{1}{2} \rho|\mathbf{v}_n|^2 
+ 
(\mathbf{J}-\rho\mathbf{v}_n)\!\cdot\!\mathbf{v}_n
+
\varepsilon_0 
\Big]d^{\,3}x
\,.
\end{equation}
The BK [1961] form of the phenomenological force $\mathbf{f}$ also implies
the dissipative heating rate, 
\begin{equation}
R 
=
(\mathbf{J}-\rho\mathbf{v}_n 
+ 
{\rm curl}\,{\lambda})\cdot{\omega}
\times (\mathbf{v}_L - \mathbf{v}_n)
\,,\nonumber\label{R-eqn-HVBK}
\end{equation}
which is Galilean invariant and positive.
Substituting this form of $\mathbf{f}$ into the superfluid motion
equation in (\ref{SF-vel-dyn}) and taking its curl provides the following
equation for the superfluid vorticity, 
${\omega}={\rm curl\,\mathbf{v}_s}$,
\begin{equation}
\partial_t\,{\omega} = {\rm curl}\,(\mathbf{v}_L\times{\omega})
\,.
\nonumber\label{omega-eqn-HVBK}
\end{equation}
This vorticity equation implies the {\bfi HVBK superfluid
Kelvin circulation theorem},
\begin{equation}
\frac{d}{dt}\int\!\!\int_{S}{\omega}\cdot d\mathbf{S} 
=
\frac{d}{dt}\oint_{\partial{S}(\mathbf{v}_L)}
\mathbf{v}_s\cdot d\mathbf{x} 
=
0
\,.\nonumber\label{circ-thm-HVBK}
\end{equation}
The Kelvin formula (\ref{circ-thm-HVBK}) expresses conservation of the
flux of superfluid vorticity through any surface $S$ whose boundary
$\partial{S}$ moves with the velocity $\mathbf{v}_L$, so $\mathbf{v}_L$ 
may be regarded as the local velocity of a vortex line. Equivalently, the
Kelvin formula expresses conservation of superfluid velocity circulation
around any loop that moves with the vortex line velocity $\mathbf{v}_L$.

In a key phenomenological step that closed the theory, BK [1961]
assigned the undetermined functions $\alpha$ and $\beta$ in the  force
$\mathbf{f}$ and its auxiliary vortex line velocity $\mathbf{v}_L$ as
\begin{equation}\label{BK-alpha-beta}
1 + \alpha\rho_s = \frac{B^{\,\prime}\rho_n}{2\rho}
\,,\quad
\beta\rho_s = \frac{B\rho_n}{2\rho}
\,.
\end{equation}
The {\it dimensionless} coefficients $B$ and $B^{\,\prime}$
were introduced earlier in Hall and Vinen (HV) [1956] to parameterize the
Gorter-Mellink [1949] ``mutual friction'' force ($B$) and its reactive
component ($B^{\,\prime}$). Hence the name, {\bfi HVBK equations} for
this closure.

The assignments in BK [1961] of the undetermined functions $\alpha$,
$\beta$ in (\ref{BK-alpha-beta}), as well as the vortex slip velocity
$\mathbf{s}^0$ in (\ref{vort-vel-zero}) are designed to reproduce the
phenomena observed in HV [1956] and yet still conserve  mass, momentum and
energy. This phenomenological approach used in BK [1961] is
indeterminate, however, in the sense that some freedom still
remains in making these assignments. The HVBK equations (\ref{SF-vel-dyn})
that result from this approach do possess the desired conservation laws
for mass, momentum and energy. And they also possess a Kelvin theorem for
the circulation of superfluid velocity. However, because of the
indeterminacy inherent in the phenomenological approach, the HVBK
equations (\ref{SF-vel-dyn}) are not unique in possessing these
properties. An alternative assignment of the vortex slip velocity is 
\begin{equation}
\mathbf{s} 
=
\mathbf{v}_\ell - \mathbf{v}_n
\quad\hbox{with}\quad
 \mathbf{v}_\ell
=
 \bar\mathbf{v} 
+ 
\rho^{-1}{\rm curl}\,{\lambda}
\,,\quad\hbox{with}\quad
\bar\mathbf{v}
=
\rho^{-1}(\rho_s \mathbf{v}_s + \rho_n \mathbf{v}_n)
\,.\label{vort-slip-vel}
\end{equation}
The mean velocity $\bar\mathbf{v}$ also figures prominantly in Hills and
Roberts [1977] discussion of the HVBK equations. As we shall see,
the alternative expression (\ref{vort-slip-vel}) for the auxiliary
vortex slip velocity in terms of the mean velocity $\bar\mathbf{v}$
arises naturally in the Hamiltonian derivation of a slightly modified set
of HVBK equations. These equations possess the same formal conservation
and circulation properties as HVBK, modulo redefining the vortex slip
velocity as $\mathbf{s}$ rather than $\mathbf{s}^0 $. The vortex slip
velocity $\mathbf{s}$ in (\ref{vort-slip-vel}) is defined relative to the
Galilean frame of the normal fluid, which is present only at finite
temperature. The HVBK $\mathbf{s}^0$ in (\ref{vort-vel-zero}) is the
limit of the vortex slip velocity $\mathbf{s}$ for zero temperature, at
which no normal fluid remains. The vortex slip velocity $\mathbf{s}$ in
(\ref{vort-slip-vel}) is a slight modification of $\mathbf{s}^0$ in
(\ref{vort-vel-zero}) necessary to incorporate finite temperature effects,
without changing the form of the HVBK theory, obtained from a Hamiltonian
derivation of these equations in the normal fluid frame. In the
Hamiltonian framework, the energy-momentum conservation laws and Kelvin
circulation theorem are all natual consequences. Moreover, the velocities
$\mathbf{v}_\ell$ and $\mathbf{v}_n$ are identified as being dual to the
momenta given by $\rho\mathbf{v}_s$ and
$\rho_n(\mathbf{v}_n-\mathbf{v}_s)$, respectively.

\paragraph{Outline.} 
We shall use a Hamiltonian approach with Lie-Poisson brackets to
derive the expression (\ref{vort-slip-vel}) for the vortex line velocity
$\mathbf{v}_\ell$ at finite temperature from first principles by using
the energy $E$ in (\ref{HVBK-erg}) as the Hamiltonian. The momenta
conjugate to the velocities $\mathbf{v}_\ell$ and $\mathbf{v}_n$ shall be
our basic dynamical variables. The finite temperature vortex line velocity
$\mathbf{v}_\ell$ and slip velocity $\mathbf{s}$ determined this way turn
out to be 
\begin{equation}
\mathbf{v}_\ell
=
\rho^{-1}(\mathbf{J} 
+ 
{\rm curl}\,{\lambda})
\,,\quad\hbox{and}\quad
\mathbf{s}
=
\mathbf{v}_\ell - \mathbf{v}_n
\simeq
\Big(\frac{\rho_s}{\rho}\Big)(\mathbf{v}_\ell^0 - \mathbf{v}_n) 
\,.\end{equation}
At zero temperature, $\rho\to\rho_s$ and these reduce to the BK [1961]
phenomenological expressions with $\mathbf{v}_\ell^0$ given by
(\ref{vort-vel-zero}). Thus, the finite temperature corrections found by
using the Hamiltonian approach {\bfi renormalize} the HVBK slip velocity 
in the mutual friction force $\mathbf{f}$ by the factor $\rho_s/\rho$ (the
superfluid mass fraction). Aside from this renormalization, the vortex line
tension is left unaffected by this renormalization, the superfluid vortex
equation keeps its form and no new parameters are added. 

Technical details of deriving this renormalized theory from its 
Hamiltonian and Lie-Poisson brackets are given in the Appendix.

\paragraph{Main results.}
We shall use the superfluid vortex dynamics for the renormalized HVBK
equations obtained via this Hamiltonian approach to write a dynamical
equation for the space-integrated total {\bfi vortex tangle length}, which
is the quantity measured in the Oregon experiments on superfluid
turbulence reported in Skrbek, Niemela and Donnelly [1999]. 

We shall also study the restriction of the renormalized HVBK equations for
the {\bfi incompressible case}, in which $\rho_n$ and $\rho_s$ are
constants and one takes $\nabla\cdot\mathbf{v}_n=0$ and
$\nabla\cdot\mathbf{v}_s=0$. Finally, we shall demonstrate the invariance
of the forms of these equations upon transforming into a rotating frame.
The Coriolis force in such a rotating frame couples to the vortex line
velocity $\mathbf{v}_L$, which of course differs from both the
superfluid velocity and the normal velocity. We shall derive a {\bfi
Taylor-Proudman theorem} for steady superfluid vortices under rapid
rotation. According to this superfluid Taylor-Proudman theorem, the vortex
line velocity becomes columnar under sufficiently rapid rotation. That is,
the lateral vortex line velocity is nondivergent and independent of the
axial coordinate, and the axial velocity decouples from the lateral
motion. Therefore,  under sufficiently rapid rotation, the superfluid
vortex filaments will straighten and become parallel to the axis of
rotation as they approach a steady state.

\paragraph{Numerical implications.} This renormalization of the
vortex line element slip velocity in the HVBK equations from
$\mathbf{s}^0\to\mathbf{s}\simeq\mathbf{s}^0\rho_s/\rho$ is sensitive to
temperature, but it does not affect the vortex line tension. Therefore, a
temperature sensitive trade-off arises between mutual friction and vortex
line tension that may be worth testing in numerical simulations such as
those reported in Henderson and Barenghi [2000].  The HVBK equations are
thought to break down in the presence of strong counterflow. However, as
general conservation laws there is no mechanism in the equations that
would signal this breakdown. A rotating Rayleigh-Besnard experiment might
be useful in testing the range of validity of the HVBK equations
(Barenghi, private communication).  Such an experiment might also
indicate how these equations should be modified in the presence of strong
counterflow.

\paragraph{Experimental implications.} Temperature sensitivity of the
coupling between the superfluid vortices and the normal fluid component is
an area of intense current investigation in superfluid Helium turbulence,
see Donnelly [1999]. One would like to know whether the $\rho_s/\rho$
renormalization of the mutual friction forces relative to the vortex line
tension would matter significantly in comparisons of the predictions of
the HVBK equations with modern experiments in Helium turbulence at low,
but finite temperatures.

\paragraph{Superfluid vortex dynamics.}
To begin addressing this issue, we may use the superfluid
vorticity equation for the renormalized HVBK equations obtained in the
Appendix via the Hamiltonian approach to write an explicit equation for
the dynamics of {\bfi Vinen's vortex length density}
$L=|\omega|/\kappa$. In the superfluid turbulence decay experiments
reported by Skrbek, Niemela and Donnelly [1999] the spatial integral of
this quantity is measured as a function of time to decrease over {\sl six
decades} as $t^{-3/2}$. The integrated vortex length measured in these
experiments is predicted by the renormalized HVBK equations to be governed by
the superfluid vorticity dynamics alone. 

Upon including mutual friction, the superfluid vortex dynamics for the
renormalized HVBK equations is expressed as, cf. equation
(\ref{omega-eqn-HVBK}),
\begin{equation}
\partial_t\,{\omega}
=
{\rm curl}(\mathbf{v}_L\times{\omega})\,,\nonumber
\end{equation}
in which the renormalized total vortex line velocity given by, cf.
equation (\ref{Phenom-force-HVBK}),
\begin{equation}\label{Total-line-vel-0}
 \mathbf{v}_L
=
\mathbf{v}_\ell
-
\frac{B^{\,\prime}\rho_n}{2\rho}\,\mathbf{s} 
-
\frac{B\rho_n}{2\rho}\,
{\hat\omega}\times\mathbf{s}
\,,\quad\hbox{where}\quad
\mathbf{s}
=
\mathbf{v}_\ell - \mathbf{v}_n
\,,
\end{equation}
and its Hamiltonian limit is found to be
\begin{equation}\label{Ideal-line-vel1}
 \mathbf{v}_\ell
=
 \bar\mathbf{v} 
+ 
\rho^{-1}{\rm curl}\,{\lambda}
\,,\quad\hbox{with}\quad
\bar\mathbf{v}
=
\rho^{-1}\mathbf{J}
\quad\hbox{and}\quad
{\lambda}
=
\lambda\,{\hat\omega}
\,.
\end{equation}
Thus, relative to the Hamiltonian approach, the terms in $B$ and $B'$ are
additional velocities introduced by phenomenology, while
$\mathbf{v}_\ell$ is the vortex line velocity in the absence of mutual
friction.

The HVBK superfluid vorticity equation implies the following dynamics for
the integrated vortex length measured in the turbulence decay experiments,
\begin{eqnarray}
\frac{d}{dt} 
\underbrace{\
\int L\, d^{\,3}x}
_{\hbox{\bfi length}}
&=& 
\underbrace{\
\int {\hat\omega}\cdot \partial_t\,{\omega}/\kappa\
d^{\,3}x
}
_{\hbox{\bfi vorticity dynamics}}  
\nonumber\\
&=&
\underbrace{\
\int L\,\tilde{\mathbf{v}}\cdot({\hat{\omega}}
\times{\rm curl}\,{\hat{\omega}})\ d^{\,3}x\
}
_{\hbox{\bfi transport $\cdot$ curvature}} 
-
\underbrace{\
\int \,L\,\frac{B\rho_n}{2\rho^2}\
({\hat{\omega}}
\times{\rm curl}\,\lambda\,{\hat{\omega}})
\cdot({\hat{\omega}}
\times{\rm curl}\,{\hat{\omega}})
\ d^{\,3}x
} 
_{\hbox{\bfi damping by curvature}}
\nonumber\\
&&+
\underbrace{\
\oint \,L\,\big(\mathbf{\hat{n}}\times{\hat\omega}\big)
\cdot
\Big(({\hat\omega}\times\mathbf{v}_\ell)
+
\frac{B\rho_n}{2\rho}(\mathbf{v}_\ell-\mathbf{v}_s)
\Big) dS
}
_{\hbox{\bfi creation and destruction at the boundary}}
\,.\label{vortex-length-dyn}
\end{eqnarray}
Here ${\hat{\omega}}$ is the unit vector tangent to a
superfluid vortex filament, so
${\kappa}=({\hat{\omega}}\times{\rm curl}\,{\hat\omega})$ 
is its local curvature. The transport and damping of the vortex tangle
length is proportional to this local curvature. The effective
velocity $\tilde{\mathbf{v}}$ in the transport term is given by 
\begin{equation}
\tilde{\mathbf{v}}
=
\overline{\mathbf{v}}
-
\frac{B^{\,\prime}\rho_n}{2\rho}\,
(\overline{\mathbf{v}} - \mathbf{v}_n)
-
\frac{B\rho_n}{2\rho}\,
{\hat{\omega}}
\times
(\overline{\mathbf{v}} - \mathbf{v}_n)
\,.\label{vortex-length-transport-vel}
\end{equation}
According to the last term in (\ref{vortex-length-dyn}), vortex length is
created or destroyed at the boundary, unless the vortex filaments approach
it in the normal direction, so that
$\mathbf{\hat{n}}\times{\hat\omega}=0$. 

Formula (\ref{vortex-length-dyn}) for the evolution of the total
superfluid vortex length presents a trade-off between the
mass-weighted velocity $\overline{\mathbf{v}}$ and the local
induction velocity (or filament curvature) ${\hat{\omega}}
\times{\rm curl}\,{\hat\omega}$, in the interior of the domain.
This trade-off in the interior competes with the process of creation and
destruction at the boundary. For example, in counterflow turbulence, the
superfluid moves toward the heater at the boundary, so the term in
$\overline{\mathbf{v}}$ would tend to be nonzero. In contrast, for grid
turbulence, $\overline{\mathbf{v}}$ is small, so this term would tend to
contribute less. This formula governs the dynamics of the experimentally
measured quantity $\int L\, d^{\,3}x$. However, it does not yet show how
to obtain the $t^{-3/2}$ decrease seen in this quantity by Skrbek, Niemela
and Donnelly [1999] in their experiments on decay of turbulence. 

Suppose the main source of decay were the term labeled ``damping by
curvature'' in formula (\ref{vortex-length-dyn}) and the flow were
isothermal and incompressible. This would imply 
\begin{equation}\label{expmnt-comparison}
\frac{1}{\langle L \rangle} \frac{d}{dt}\langle L \rangle
=
-\,c_0(T)\, \lambda_0 \, \frac{\langle   L/R^2 \rangle}{\langle L \rangle}
=
-\  \frac{3}{2\,(t+t_0)}
\,,
\end{equation}
where $\lambda_0=\lambda/\rho_s = (\kappa/4\pi) ln (b/a)$ is the quantum
vortex constant, $t_0$ is a time shift in the experimental analysis,
$c_0(T)\equiv B\rho_n\rho_s/(2\rho^2)$ and angle brackets
$\langle{\cdot}\rangle$ denote spatial integral over the
measurement domain. In particular,
\begin{equation}
\langle L \rangle
\equiv
\int L\, d^{\,3}x
\,,\quad
\langle   L/R^2 \rangle
\equiv
\int L\ |{\hat{\omega}}
\times{\rm curl}\,{\hat{\omega}}|^2\, d^{\,3}x
\,.
\end{equation}
The measured $t^{-3/2}$ decrease in $\langle L\rangle$ implies via
formula (\ref{expmnt-comparison}) that the length-weighted mean curvature
of the vortex tangle $\langle L/R^2\rangle / \langle L\rangle$ decays due
to mutual friction as $t^{-1}$. Thus, on the average {\bfi as the
vortex length decays, the vortices tend to straighten}, under the effects
of mutual friction damping.

\paragraph{Preservation of helicity versus preservation of vortex
length.}  The {\bfi helicity, or linkage number} for the superfluid 
vorticity is defined as
\begin{equation}
 \Lambda
=
 \int (\mathbf{v}_s\cdot{\omega})\ d^{\,3}x
\,.\nonumber
\end{equation}
The helicity satisfies an evolution equation obtained from the superfluid
vortex dynamics,
\begin{equation}
 \frac{d\Lambda}{dt}
=
 -
\oint (\mathbf{\hat{n}}\cdot{\omega})
\Big(\mu - \frac{1}{2}\,v_n^2 - 
\mathbf{v}_s\cdot(\mathbf{v}_L -\mathbf{v}_n)\Big)\
dS
 -
\oint (\mathbf{\hat{n}}\cdot\mathbf{v}_L)
(\mathbf{v}_s\cdot{\omega})\ dS
\,.\nonumber
\end{equation}
Therefore, even with mutual friction, helicity is created and destroyed
only on the boundary. Moreover, helicity will be preserved, provided both
${\omega}$ and $\mathbf{v}_L$ are {\sl tangential at the
boundary}. The former condition, however, is the opposite of that required
for the creation and destruction of vortex length at the boundary to
cease. Therefore, no equilibrium should be expected that preserves both
the helicity and the vortex length in a superfluid.
\smallskip

\paragraph{Superfluid vortex equilibria are not ABC flows.}
The steady equilibrium solutions of the superfluid vorticity
dynamics satisfy
\begin{equation}
{\rm curl}\,(\mathbf{v}_L\times{\omega})
=0\,.
\label{super-Beltrami}
\end{equation}
For example, a steady equilibrium exists when ${\omega}$ and
$\mathbf{v}_L$ are parallel. Note that these ``super-Beltrami flows'' are
{\sl not} eigenfunctions of the curl. Therefore, they are not
Arnold-Beltrami-Childress (ABC) flows, as occur for the Euler equations.

\section{Incompressible renormalized HVBK flows}

To express the renormalized HVBK equations in the incompressible limit, we
begin by recollecting the compressible equations and abbreviating
$|\mathbf{v}_s - \mathbf{v}_n|^2=v_{s\,n}^2\,$, 
\begin{eqnarray}
\partial_t S
&=&
-\,{\rm div}(S\mathbf{v}_n)
+
R/T
\,,\nonumber\\
\partial_t\rho
&=&
-\,{\rm div}(\rho_s \mathbf{v}_s + \rho_n \mathbf{v}_n)
\,,\nonumber\\
\rho_s\big(\partial_t \mathbf{v}_s
+
(\mathbf{v}_s\cdot\nabla)\mathbf{v}_s\big)
&=&
-\,\rho_s
\nabla\big(\,\mu 
- \frac{1}{2}v_{s\,n}^2\, \big)
+
\rho_s(\mathbf{v}_L-\mathbf{v}_s)\times {\omega}
\,,\nonumber\\
\partial_t \big(\rho_s {v}_{s\,i } + \rho_n {v}_{n\,i }\big)
&=&
-\,
\partial_k \, \big(\rho_n {v}_{n\,i }{v}_n^k 
+
\rho_s {v}_{s\,i } {v}_s^k\big)
-\,
\partial_i P
-\,
\partial_k\tau_i^k
\,,\nonumber\\
\tau_i^k
&=&
\epsilon^{klm}\,\partial_{\,l} \big({v}_{s\,i\,}\lambda_m
\big)
-
\lambda_i\,\omega^k
+
\delta_i^k\,{\lambda\cdot\omega}
\,.\nonumber
\end{eqnarray}
As we have seen, finite temperature effects renormalize the total vortex
line velocity as
\begin{equation}\label{Total-line-vel}
 \mathbf{v}_L
=
\mathbf{v}_\ell
-
\frac{\rho_n}{\rho}\bigg(
\frac{B}{2}\,
{\hat\omega}\times\mathbf{s}
+
\frac{B^{\,\prime}}{2}\,\mathbf{s} 
\bigg)
\,,\quad\hbox{where}\quad
\mathbf{s}
=
\mathbf{v}_\ell - \mathbf{v}_n
\,,
\end{equation}
and the Hamiltonian part of the line velocity 
(with corrections for finite temperature) is defined as
\begin{equation}\label{Ideal-line-vel2}
 \mathbf{v}_\ell
=
 \bar\mathbf{v} 
+ 
\rho^{-1}{\rm curl}\,{\lambda}
\,,\quad\hbox{with}\quad
\bar\mathbf{v}
=
\rho^{-1}(\rho_s \mathbf{v}_s + \rho_n \mathbf{v}_n)
\quad\hbox{and}\quad
{\lambda}
=
\lambda\,{\hat\omega}
\,.
\end{equation}

To the extent that $\rho$, $\rho_s$, $\rho_n$ and $S$ all may be taken as
constants for a given temperature and the heating rate $R$ is negligible,
then the velocities $\mathbf{v}_n$ and $\mathbf{v}_s$ are incompressible,
i.e.,
\begin{equation}
\nabla\cdot\mathbf{v}_n=0
\quad\hbox{and}\quad
\nabla\cdot\mathbf{v}_s=0
\,.\nonumber
\end{equation}
In this situation, the pressure $P$ may be obtained from the {\bfi
Poisson equation},
\begin{equation}\label{Pressure-eqn}
- \nabla^2 \big( P + {\lambda\cdot\omega} \big)
=
{\rm div}\Big(\rho_s(\mathbf{v}_s\cdot\nabla)\,\mathbf{v}_s
+
\rho_n(\mathbf{v}_n\cdot\nabla)\,\mathbf{v}_n
\,-\,
{\omega}\cdot\nabla{\lambda}\Big)
\,,
\end{equation}
found from the divergence of the total momentum equation. Combining the
superfluid motion equation with total momentum conservation results in 
an equation for the normal fluid velocity in the incompressible case
\begin{equation}
\rho_n\partial_t \mathbf{v}_n
+
\rho_n(\mathbf{v}_n\cdot\nabla)\mathbf{v}_n
=-\,
\nabla\big(\,P^{\,\prime} - \rho_s\mu 
+ 
\frac{\rho_s}{2}v_{s\,n}^2\,  \big)
-
\rho_s (\mathbf{v}_L-\mathbf{v}_s)\times {\omega}
+
{\omega}\cdot\nabla{\lambda}
\,,\nonumber
\end{equation}
where $\mathbf{v}_L$ is given in equation (\ref{Total-line-vel}). We
set $P^{\,\prime} \equiv P+{\lambda\cdot\omega}$ and take it as the total 
pressure.  (One also could have absorbed ${\lambda\cdot\omega}$ into $P$
earlier, by including it in Euler's pressure law.) Since
${\lambda}=|\lambda|{\hat\omega}$ and 
${\hat\omega}$ is a unit vector, we find {\sl for constant}
$\rho_s$ the standard relation for the {\bfi vortex line tension} denoted
as $\mathbf{T}$. Namely,
\begin{equation}
{\omega}\cdot\nabla{\lambda}
=
-\, \lambda_0\,\rho_s\,{\omega}
\times{\rm curl}\,{\hat\omega}
\equiv
\rho_s\mathbf{T}
\,,\nonumber
\end{equation}
where $\lambda_0=\lambda/\rho_s = (\kappa/4\pi){\rm ln} (b/a)$ is a
constant. 

\noindent
{\bf Remark.} We note that the quantity $\mathbf{T}$ known as the vortex
line tension first appears in the {\sl normal fluid equation}, as a
reaction to the presence of the superfluid. The standard convention for
introducing the mutual friction force has the effect of shifting
$\mathbf{T}$ into the superfluid equation. By action and reaction, though,
$\mathbf{T}$ could appear in either equation.

These equations of motion must be completed by providing an equation of
state relation for the quantity $\mu  - \frac{1}{2}v_{s\,n}^2$. 
BK [1961] assumes a law of partial pressures,
\begin{equation}
P_n = \frac{\rho_n}{\rho}\,P^{\,\prime}  = P^{\,\prime}  - P_s
\quad\hbox{and}\quad
P_s = \frac{\rho_s}{\rho}\,P^{\,\prime} 
=
\rho_s\,\mu 
- 
\frac{1}{2}\rho_s \, v_{s\,n}^2 
\,.\nonumber
\end{equation}
In this case, the renormalized HVBK motion equations for incompressible
flow reduce to
\begin{eqnarray}\label{incomp-super-motion}
\partial_t \mathbf{v}_s
+
(\mathbf{v}_s\cdot\nabla)\mathbf{v}_s
&=&
-\,\frac{1}{\rho}\,
\nabla P^{\,\prime} 
-
\frac{\rho_n}{\rho}
\,\mathbf{F}_{n\,s}
+
\mathbf{T}
\,,\\
\partial_t \mathbf{v}_n
+
(\mathbf{v}_n\cdot\nabla)\mathbf{v}_n
&=&-\,\frac{1}{\rho}
\nabla P^{\,\prime} 
+
\frac{\rho_s}{\rho}
\,\mathbf{F}_{n\,s}
\,.\label{incomp-normal-motion}
\end{eqnarray}
In these superfluid motion equations, the {\bfi renormalized mutual
friction force} $\mathbf{F}_{n\,s}$ is defined as the sum (with
${\omega}={\rm curl}\,\mathbf{v}_s$)
\begin{equation}
\mathbf{F}_{n\,s}
=
(\mathbf{s}^{\,0}\times {\omega})
+
\Big(\frac{\rho_s}{\rho}\Big)\,
\,\mathbf{F}_{n\,s}^{\,0}
\,,
\quad\hbox{where}\quad
\mathbf{F}_{n\,s}^{\,0}
=
\Big( 
 \frac{B}{2}\,{\hat\omega}\times\mathbf{s}^{\,0}
+
 \frac{B^{\,\prime}}{2}\,\mathbf{s}^{\,0} 
\Big)\times {\omega}
\,.
\label{F-NS-new}
\end{equation}
Here $\mathbf{F}_{n\,s}^{\,0}$ is the HVBK mutual friction force without
any finite temperature corrections. To acquire these formulas, we used the
relations for the incompressible case, 
\begin{equation}\label{slip-vel}
\mathbf{s}
=
\mathbf{v}_\ell - \mathbf{v}_n
=
\frac{\rho_s}{\rho}\big(\mathbf{v}_s 
+ 
\lambda_0\,{\rm curl}\,{\hat\omega}
-
\mathbf{v}_n
\big)
=
\frac{\rho_s}{\rho}\big(\mathbf{v}_\ell^0 - \mathbf{v}_n\big)
=
\frac{\rho_s}{\rho}\,\mathbf{s}^{\,0}
\,,
\end{equation}
with
$\mathbf{s}^{\,0}
=
\mathbf{v}_s 
+ 
\lambda_0\,{\rm curl}\,{\hat\omega}
-
\mathbf{v}_n\,,
$
and we eliminated $\mathbf{v}_L$ by using the relation
\begin{equation}\label{dissip-vortex-vel}
-\,\rho_s\big(\mathbf{v}_L - \mathbf{v}_s
-\lambda_0\,{\rm curl}\,{\hat\omega}\big)
=
\rho_n\,\mathbf{s}
+ 
\frac{\rho_s\rho_n}{\rho}
\bigg( 
 \frac{B}{2}\,{\hat\omega}\times\mathbf{s}
+
 \frac{B^{\,\prime}}{2}\,\mathbf{s} 
\bigg)\,.
\end{equation}
In equation (\ref{F-NS-new}) for $\mathbf{F}_{n\,s}$, the quantity
$(\mathbf{s}^{\,0}\times{\omega})$ is the {\bfi Hamiltonian
reactive force} (which could  be naturally absorbed into Vinen's
$B^{\,\prime}$ parameter) and $\mathbf{F}_{n\,s}^{\,0}$ is the
phenomenological mutual friction force defined according to the standard
convention as in BK [1961] and    Donnelly [1999]. The finite-temperature
corrections contribute an {\bfi overall factor} of $\rho_s/\rho$ to the
standard zero-temperature expression $\mathbf{F}_{n\,s}^{\,0}$ for the
phenomenological mutual friction force. No new parameters are added, but a
{\bfi temperature dependent trade-off} is identified between the
renormalized mutual friction coupling and the vortex line tension, since
the vortex line tension remains {\sl unaffected} by the finite-temperature
corrections. 

In the isothermal case, the motion equations are closed by the Poisson
equation for $P$, since the other coefficients ($B$, $B^{\,\prime}$,
$\rho_n/\rho$, etc.)  are specified functions of temperature and they may
be taken as constants, for an isothermal incompressible superfluid flow. 

Note that equations
(\ref{incomp-super-motion}-\ref{incomp-normal-motion})  may be rewritten
with ${\omega}_n={\rm curl}\,\mathbf{v}_n$ as
\begin{eqnarray}\label{incomp-super-motion2}
\partial_t \mathbf{v}_s
+
\nabla\mu_s
&=&
\mathbf{v}_n\times{\omega}
+
\frac{\rho_s}{\rho}\mathbf{s}^{\,0}
\times{\omega}
-
\frac{\rho_n\rho_s}{\rho^2}\,
\mathbf{F}_{n\,s}^{\,0}
\quad\hbox{with}\quad
\mu_s=P^{\,\prime}\!/\rho + \frac{1}{2}v_s^2
\,,\\
\partial_t \mathbf{v}_n
+
\nabla\mu_n
&=&
\mathbf{v}_n\times{\omega}_n
+
\frac{\rho_s}{\rho}\mathbf{s}^{\,0}
\times{\omega}
+
\frac{\rho_s^2}{\rho^2}\,
\mathbf{F}_{n\,s}^{\,0}
\quad\hbox{with}\quad
\mu_n=P^{\,\prime}\!/\rho + \frac{1}{2}v_n^2
\,.\label{incomp-normal-motion2}
\end{eqnarray}
These equations imply an equation for the {\bfi velocity difference},
\begin{equation}\label{velocity-diff}
\partial_t (\mathbf{v}_s-\mathbf{v}_n)
+
\frac{1}{2}\nabla(v_s^2-v_n^2)
=
\mathbf{v}_n\times
({\omega}-{\omega}_n)
-
\frac{\rho_s}{\rho}
\mathbf{F}_{n\,s}^{\,0}
\quad\hbox{with}\quad
{\omega}_n
=
{\rm curl}\,\mathbf{v}_n
\,.
\end{equation}
and there is {\sl no tendency} for mutual friction to cause any 
alignment in the vorticities of the superfluid and its normal component. 
Instead, the curl$\,{\hat\omega}$ part of $\,\mathbf{F}_{n\,s}^{\,0}\ne0$
would break any such alignment, if it were to form spontaneously. Indeed,
{\bfi alignments \underline{sufficient} for steady solutions} are
\begin{equation}\label{vortex-align}
\mathbf{s}^{\,0}\times (\nabla\mu_s\times\nabla\mu_n)
=
0
\,,\
\mathbf{s}^{\,0}\times {\omega} = 0
\
\hbox{and}\
\mathbf{s}^{\,0}\times\mathbf{v}_n
=
0
\,,
\quad\hbox{with}\quad
\mathbf{s}^{\,0}
\equiv
\mathbf{v}_s 
+ 
\lambda_0\,{\rm curl}\,{\hat\omega}
-
\mathbf{v}_n
\,,\nonumber
\end{equation}
provided $\mu_s$ and $\mu_n$ are functionally unrelated. Thus, the steady
equilibrium alignments imposed by mutual friction involve $\mathbf{v}_n$,
$\mathbf{v}_s$ and curl$\,{\hat\omega}$, as well as the
independent gradients of $\mu_s$ and $\mu_n$. For example, one class of
equilibria has $\mathbf{s}^{\,0}$, $\mathbf{v}_n$, ${\omega}$ all aligned
tangent to intersections of level surfaces of $\mu_n$ and $\mu_s$.

\section{Rotating frame renormalized HVBK equations}

We transform to a rotating frame with relative velocities denoted with
an asterisk as $\mathbf{v}_s^*=\mathbf{v}_s-\mathbf{R}(\mathbf{x})$, etc.,
and curl$\,\mathbf{R}=2{\Omega}$. After a calculation involving
Legendre transformations, we obtain the Hamiltonian for the relative
motion, cf. the Hamiltonian in (\ref{SF-Ham-def}) of the Appendix,
\begin{eqnarray}\label{Ham-rot}
h (\mathbf{M}^*, \rho, S, \mathbf{u}, \mathbf{A}^*,  n) 
&=&
\int \bigg\{ \frac{1}{2} 
\rho\, |\mathbf{v}_n^* + \mathbf{R}(\mathbf{x})|^2 
+
(\mathbf{M}^* - \rho\mathbf{A}^* - \rho\mathbf{v}_n^*)\cdot\mathbf{v}_n^*
\\
&&\hspace{-2cm}+\
\varepsilon_0 (\rho,S,\mathbf{v}_s^*-\mathbf{v}_n^*, 
\,{\omega}^* + 2{\Omega})
-
\mathbf{R}\cdot\Big[\rho(\mathbf{v}_n^* + \mathbf{R})
+
(\rho-n)(\mathbf{A}^* - \mathbf{R})\Big]
\bigg\} d^{\,3}x
\,.\nonumber
\end{eqnarray}
Here $\mathbf{M}^* - \rho\mathbf{A}^* =
\mathbf{J}^*=\mathbf{J}-\rho\mathbf{R}$ and 
$\mathbf{v}_s = \mathbf{u} - (\mathbf{A}^*-\mathbf{R})$. The equations
resulting from the Lie-Poisson bracket (\ref{semidirect-LPB-2})  of the
Appendix in these relative variables keep their forms and the
condition
$n=\rho$ is still preserved. We conclude with the following three remarks.
\paragraph{Superfluid Coriolis force couples to the vortex line
velocity.} The Hamiltonian evolution equation for the superfluid velocity
in the rotating frame is expressed as 
\begin{equation}
\partial_t \mathbf{v}_s^*
+
(\mathbf{v}_s^*\cdot\nabla)\mathbf{v}_s^*
=
-\,
\nabla\big(\,\mu 
- \frac{1}{2}|\mathbf{v}_s^* - \mathbf{v}_n^*|^2 
- \frac{1}{2}|\mathbf{R}|^2\big)
+
 (\mathbf{v}_\ell^*-\mathbf{v}_s^*)\times {\omega}^*
+
 \mathbf{v}_\ell^*\times 2{\Omega}
\,.\nonumber
\end{equation}
The last term is the Coriolis force and it involves the relative vortex
line velocity. The curl of this equation yields
\begin{equation}
\partial_t ({\omega}^* + 2{\Omega})
=
{\rm curl}\
\big(\mathbf{v}_\ell^*\times
({\omega}^* + 2{\Omega})\big)
\,.\nonumber
\end{equation}
The form of the vortex dynamics equation is {\bfi invariant}
under passing to a steadily rotating frame, and the superfluid Coriolis
force contains the renormalized vortex line velocity, rather than the
superfluid velocity. Therefore, this is not merely a kinematic force. The
vortex line velocity appearing in the superfluid Coriolis force
includes the interaction between the vortex lines and the superfluid
component. It also includes the interaction with the normal component,
since $\mathbf{v}_\ell$ depends on the relative momentum density and
contains finite temperature effects. The superfluid Coriolis force is
essential in the spin up problem in He-II, see, e.g., Reissenegger [1993].

\paragraph{Superfluid Taylor-Proudman theorem.} For steady, or slow
motions and rapid rotation we have
\begin{equation}
0
=
{\rm curl}\
\big(\mathbf{v}_\ell^*\times
2{\Omega}\big)
\,.\nonumber
\end{equation}
If the rotation is uniform ($\nabla{\Omega}=0$)
and oriented vertically (${\Omega} = |\Omega|\mathbf{\hat z}$)
this becomes
\begin{equation}
0
=
2|\Omega|\
\big(\partial_z\,\mathbf{v}_\ell^*
- 
\mathbf{\hat z}\,{\rm div}\,\mathbf{v}_\ell^*\big)
=
2|\Omega|\
\big(\partial_z\,{v}_{\ell\,x}^*,\,
\partial_z\,{v}_{\ell\,y}^*,\,
- 
\partial_x\,{v}_{\ell\,x}^*
-
\partial_y\,{v}_{\ell\,y}^*\big)^T
\,,\nonumber
\end{equation}
where $(\ )^T$ denotes transpose of a row vector into a column vector.
Thus, for steady, or slow motions and rapid uniform rotation, we find that
{\bfi vortex line motion is columnar}. That is, the lateral
vortex line velocity is nondivergent and independent of the axial
coordinate, and the axial velocity decouples from the lateral motion.
Therefore,  under sufficiently rapid rotation, the superfluid vortex
filaments will straighten and become parallel to the axis of rotation as
they approach a steady state. However, they may still undergo nondivergent
motion in the lateral plane. This superfluid Taylor-Proudman theorem
explains why steady superfluid vortices tend to be aligned with the
rotation axis under rapid uniform rotation.  The same conclusion applies,
if the velocity $\mathbf{v}_\ell^*$ in the Hamiltonian formulation is
replaced by the phenomenological relative velocity
$\mathbf{v}_L^*=\mathbf{v}_L-\mathbf{R}$. Similar considerations are
discussed in Sonin [1987] from a more microscopic viewpoint.

\paragraph{Relative total momentum is not conserved for rotating 
compressible flows.} Since the Hamiltonian depends explicitly on spatial
position, instead of conserving relative total momentum, we have the
balance
\begin{equation}
\partial_t {J}_i^* + \partial_j T_i^{\,*\,j}
=-\
\frac{\partial\, \mathsf{h}}{\partial x^i}\Big|_{explicit}
=
\frac{\rho}{2}\,\partial_i|\mathbf{R}|^2
\,,\nonumber
\end{equation}
where $\mathsf{h}$ is the Hamiltonian density in equation (\ref{Ham-rot}).
This relative momentum balance is the effect of centrifugal force. Here we
have dropped terms proportional to $\rho-n$, since $\rho=n$ is still 
preserved in a rotating frame. Consequently, the stress tensor in the
relative momentum equation also keeps its form in passing to a rotating
frame, although the total relative momentum is no longer conserved if the
flow is compressible.

\section*{Acknowledgments}
I am grateful to H. R. Brand, A. Brandenburg, P. Constantin, R. Donnelly,
V. V. Lebedev, F. Lund, J. E. Marsden, J. Niemela, A. Reisenegger, L.
Skrbek, K. Sreenivasan and W. F. Vinen for stimulating discussions and
encouragement.  I am also grateful for hospitality at the UC Santa Barbara
Institute for Theoretical Physics where this work was initiated during
their Hydrodynamic Turbulence program in spring 2000. This research was
supported by the U.S. Department of Energy under contracts W-7405-ENG-36
and the Applied Mathematical Sciences Program KC-07-01-01.

\section*{Appendix: Lie-Poisson Hamiltonian formulation}

Conservation of the number of quantum vortices moving through superfluid
$^4$He (and across the streamlines of the normal fluid component) is
expressed by
\begin{equation}
\frac{d}{dt} \int\!\!\int_{S}
{\omega\cdot\hat n}\,dS
=
0
\,,\label{vort-cons}
\end{equation}
where the superfluid vorticity ${\omega}$ is the areal density
of vortices and ${\hat n}$ is the unit vector normal to the
surface $S$ whose boundary $\partial S$ moves with the vortex line
velocity $\mathbf{v}_\ell$. When 
${\omega}={\rm curl}\,\mathbf{v}_s$ this is
equivalent to a {\bfi vortex Kelvin theorem}
\begin{equation}
\frac{d}{dt} \oint_{\partial S(\mathbf{v}_\ell)}
\mathbf{v}_s\cdot d\mathbf{x}
=
0
\,,\label{vort-Kel}
\end{equation}
which in turn implies the fundamental relation
\begin{equation}
\partial_t \mathbf{v}_s
-
\mathbf{v}_\ell\times{\omega}
=
\nabla\mu
\,.\label{phase-slip}
\end{equation}

The superfluid velocity naturally splits into 
$\mathbf{v}_s = \mathbf{u} - \mathbf{A}$, where $\mathbf{u} = \nabla\phi$
and (minus) the curl of $\mathbf{A}$ yields the superfluid vorticity
${\omega}$. The phase $\phi$ is then a regular function without
singularities.  This splitting will reveal that the Hamiltonian dynamics
of superfluid $^4$He with vortices may be expressed as an invariant
subsystem of a larger Hamiltonian system in which $\mathbf{u}$ and
$\mathbf{A}$ have independent evolution equations. 

We begin by defining a phase frequency in the normal velocity frame as 
\begin{equation}
\partial_t\phi + \mathbf{v}_n\cdot\nabla \phi
=
\nu
\,.
\end{equation}
The mass density $\rho$ and the phase $\phi$ are canonically
conjugate in the Hamiltonian formulation. Therefore, one may set 
$\nu = - \,\delta h/\delta \rho$ for a given Hamiltonian $h$ and
the phase gradient $\mathbf{u}=\nabla\phi$ satisfies
\begin{equation}
\partial_t\mathbf{u} + \mathbf{v}_n\cdot\nabla\mathbf{u}
+ (\nabla\mathbf{v}_n)^T\cdot \mathbf{u}
=
- \,\nabla\frac{\delta h}{\delta \rho}
\,,
\end{equation}
where $(\ )^T$ denotes transpose, so that $(\nabla\mathbf{v}_n)^T\cdot
\mathbf{u} = u_j\nabla v_n^j$. The mass density $\rho$ satisfies the dual
equation
\begin{equation}
\partial_t\rho + \nabla \cdot(\rho\mathbf{v}_n)
=
- \,\nabla\cdot \frac{\delta h}{\delta \mathbf{u}}
\,.
\end{equation}
Perhaps not surprisingly, the rotational and potential components of the
superfluid velocity $\mathbf{v}_s=\mathbf{u}-\mathbf{A}$ satisfy similar
equations, but the rotational component is advected by the vortex line
velocity
$\mathbf{v}_\ell$, instead of the normal velocity $\mathbf{v}_n$.
Absorbing all gradients into $\mathbf{u}$ yields
\begin{equation}
\partial_t \mathbf{A}
+
\mathbf{v}_\ell\times{\omega}
=
0
\,.\label{A-eqn}
\end{equation}
Taking the difference of the equations for $\mathbf{u}$ and
$\mathbf{A}$ then recovers equation (\ref{phase-slip}) as
\begin{equation}
\partial_t \mathbf{v}_s
-
\mathbf{v}_\ell\times{\omega}
=
- \,\nabla\Big(\mathbf{v}_n\cdot \mathbf{u} + \frac{\delta h}{\delta
\rho}\Big)
\quad\hbox{with}\quad
\mathbf{v}_s = \mathbf{u} - \mathbf{A}\label{vee-eqn}\,,
\end{equation}
in which one uses regularity of the phase $\phi$ to set
curl$\,\mathbf{u}=0$. It remains to determine $\mathbf{v}_\ell$ from
the Hamiltonian formulation. Including the additional
degree of freedom $\mathbf{A}$  associated with the vortex lines allows
them to move relative to both the normal and super components of the
fluid, and thereby introduces an additional reactive force without
introducing any additional inertia. This Hamiltonian approach thus yields
{\bfi renormalized HVBK equations}.

\paragraph{Proposition:} 
{\it Upon splitting the superfluid velocity into 
$\mathbf{v}_s = \mathbf{u} - \mathbf{A}$ (with $\mathbf{u} = \nabla\phi$
so that ${\omega} = -\,{\rm curl}\,\mathbf{A}$) the (renormalized)
HVBK equations in the Galilean frame of the normal fluid form an 
{\bfi invariant subsystem of a Lie-Poisson Hamiltonian system},
\begin{equation}
\frac{\partial{f}}{\partial{t}} = \{f,h\}
\quad\hbox{with}\quad
f,h\in(\mathbf{M},\rho,S,\mathbf{u},\mathbf{A},n)
\,,\nonumber
\end{equation}
and {\bfi Lie-Poisson bracket} given by
}
\begin{eqnarray}
\{f,h\}
&=&-\int
\bigg\{  
\frac{\delta{f}}{\delta{{M}_j}}
\bigg[
({M}_k\partial_j + \partial_k{M}_j)
\frac{\delta{h}}{\delta{{M}_k}}
+ 
\rho\,\partial_j\frac{\delta{h}}{\delta{\rho}}
+ 
{S}\partial_j\frac{\delta{h}}{\delta{{S}}}
+\
(\partial_k{u}_j - {u}_{k\,,\,j})
 \frac{\delta{h}}{\delta{{u}_k}}
\bigg]
\nonumber\\
&&
+\,
\bigg[  
\frac{\delta{f}}{\delta{\rho}}
\partial_k \rho
+\,
\frac{\delta{f}}{\delta{S}}
\partial_k\, {S}
+\,
\frac{\delta{f}}{\delta{u_j}}
(u_k\partial_j + u_{j\,,\,k})
\bigg]\frac{\delta h}{\delta{M}_k}
+\,
\bigg[  
\frac{\delta{f}}{\delta{\rho}}
\partial_k
\frac{\delta{h}}{\delta{{u}_k}}
+
\frac{\delta{f}}{\delta{{u}_j}}
\partial_j\frac{\delta{h}}{\delta{\rho}}
\bigg]
\nonumber\\
&&\hspace{15mm}
-\  
\frac{\delta{f}}{\delta{{A}_j}}
\bigg[
\partial_j \frac{\delta{h}}{\delta{n}}
+
\frac{A_{j\,,\,k} - A_{k\,,\,j}}{n}
\frac{\delta{h}}{\delta{{A}_k}}
\bigg]
-
\frac{\delta{f}}{\delta{n}}
\partial_k
 \frac{\delta{h}}{\delta{A_k}}
\bigg\} d^{\,3}x
\,.\label{semidirect-LPB-2}
\end{eqnarray}
%
\paragraph{Remarks:} Here $\mathbf{M}$ is the total momentum density, the
total mass density is $\rho$ and the entropy density is $S$. We shall
interpret the density $n$ later, after we develop the Hamiltonian
equations of motion. It shall emerge that $n=\rho$ is an invariant
condition and, hence, 
\begin{equation}
\mathbf{M} - n\mathbf{A}
=
\mathbf{J}
=
\mathbf{p} + \rho\mathbf{v}_n
\,,\nonumber
\end{equation}
for $n=\rho$, where
$
\mathbf{p} = \mathbf{J} - \rho\mathbf{v}_n 
= \rho_s (\mathbf{v}_s - \mathbf{v}_n)
$ 
is the relative momentum density of the superfluid in the Galilean frame
of the normal fluid. The momentum density associated with the vortex fluid
will be $\mathbf{N}= - n\mathbf{A}$. The Hamiltonian will be the energy 
$E$ in (\ref{HVBK-erg}).

The Lie-Poisson bracket in the Proposition appeared first in Holm and
Kupershmidt [1987] in a study of various approximate equations for the
dynamics of multicomponent superfluids with charged condensates. The
mathematical structure of this Lie-Poisson bracket and its association
with the dual of a certain Lie algebra is discussed in Holm and
Kupershmidt [1987]. Our re-interpretation of this Poisson bracket 
introduced and studied earlier shall now yield a extension of the HVBK
equations that enables the vortex line velocity $\mathbf{v}_\ell$ and
hence the vortex reactive force and mutual friction force to be expressed
at finite temperature. Identifying this Poisson bracket as being dual to a
Lie algebra establishes that it satisfies the Jacobi identity,
$\epsilon_{ijk}\{f_i\,,\{f_j\,,f_k\}\}=0$. The term in the Poisson bracket
responsible for the reactive vortex force will turn out to be
$\{A_i\,,\,A_j\}\ne0$. The Poisson bracket
$\{v_{s\,i}\,,\,v_{s\,j}\}$ would vanish (as does 
$\{u_i\,,\,u_j\}=0$) and thus the reactive vortex force would be absent,
in any Hamiltonian formulation for which $\mathbf{A}$ and $n$ were not
independent degrees of freedom from $\mathbf{M}$, $\rho$, $S$. Volovik and
Dotsenko [1979, 1980] obtain a different result and provide no
Lie-algebraic justification.

A Lagrangian formulation of these equations is also available. However,
it involves an equation for $\delta l/\delta \nu$ about which nothing is
known physically.

\paragraph{Corollary \#1:} 
{\it The Lie-Poisson bracket is equivalent to the
following separate Hamiltonian matrix forms for the dynamical equations
\begin{equation}\label{Ham-matrix-SF1}
\frac{\partial}{\partial t}
\left[ \begin{array}{c} 
M_i \\ S \\ \rho \\ u_i 
\end{array}\right]
= -
\left[ \begin{array}{cccc} 
M_j\partial_i + \partial_j M_i & 
S  \partial_i & \rho\partial_i &
 \partial_j u_i - u_{j\,,\,i}  
\\ 
\partial_j S & 0 & 0 & 0
\\
\partial_j\rho & 0 & 0 & \partial_j 
\\
u_j\partial_i + u_{i\,,\,j}& 0 & \partial_i & 0 
\end{array} \right]
\left[ \begin{array}{c} 
{\delta h/\delta M_j} \\ 
{\delta h/\delta S} \\ 
{\delta h/\delta \rho} \\ 
{\delta h/\delta u_{\,j}} 
\end{array}\right],
\end{equation}
and, upon defining $\mathbf{N} = - n \mathbf{A}$,
\begin{equation}\label{Ham-matrix-SF2}
\frac{\partial}{\partial t}
\left[ \begin{array}{c} 
N_i \\ n  
\end{array}\right]
= -
\left[ \begin{array}{cc} 
N_j\partial_i + \partial_j N_i & 
n\partial_i 
\\ 
\partial_jn & 0 
\end{array} \right]
\left[ \begin{array}{c} 
{\delta h/\delta N_j} \\ 
{\delta h/\delta n} 
\end{array}\right].
\end{equation}
These are individually expressed as
%
\begin{eqnarray}
\partial_t S
&=&
\{S,h\}
=
-\,{\rm div}(S\, {\delta h/\delta \mathbf{M}})
\,,\nonumber\\
\partial_t n
&=&
\{n,h\}
=
-\,{\rm div}(n \, {\delta h/\delta \mathbf{N}})
\,,
\nonumber\\
\partial_t\rho
&=&
\{\rho,h\}
=
-\,{\rm div}(\rho \, {\delta h/\delta \mathbf{M}} 
+ {\delta h/\delta \mathbf{u}})
\,,\nonumber\\
\partial_t \mathbf{u}
&=&
\{\mathbf{u},h\}
=
-\,
\nabla\big(\, {\delta h/\delta \rho 
+ 
(\delta h/\delta \mathbf{M}})\cdot\mathbf{u}\big)
+
(\delta h/\delta \mathbf{M})\times {\rm curl}\,\mathbf{u}
\,,\nonumber\\
\partial_t (\mathbf{N}/n)
&=&
\{(\mathbf{N}/n),h\}
=
-\,
\nabla\big(\, \delta h/\delta n 
+ 
(\delta h/\delta \mathbf{N})\cdot(\mathbf{N}/n)\big)
+
\big(\delta h/\delta \mathbf{N}\big)
\times {\rm curl}\,(\mathbf{N}/n)
\,,\nonumber\\
\partial_t \big(M_j + N_j \big)
&=&
\{M_j + N_j,h\}
=
-\,
\partial_k \, T_j^{\,k}
\,.\nonumber
\end{eqnarray}
}

\paragraph{Corollary \#2:} 
{\it Consider a translation invariant Hamiltonian density with dependence
\[h(\mathbf{M},\rho,S,n,\mathbf{v}_s,{\omega},\mathbf{A}),\]
where $\mathbf{v}_s=\mathbf{u}-\mathbf{A}$, $\mathbf{A}=-\mathbf{N}/n$
and ${\omega}={\rm curl}\,\mathbf{v}_s$. The stress
tensor $T_j^{\,k}$ is expressed in terms of this Hamiltonian as
\begin{equation} \label{2nd-stress-tensor}
T_j^{\,k} 
= 
M_j\frac{\partial h}{\partial M_k}
+
v_{s\,j}\bigg(\frac{\partial h}{\partial v_{s\,k}}
+
\Big({\rm curl}\,\frac{\partial h}{\partial {\omega}}\Big)_k
\bigg)
-
v_{s\,l,\,j}\epsilon_{mlk}\,\frac{\partial h}{\partial \omega_m}
+\
\delta_j^{\,k} P
-
A_j\,\frac{\partial h}{\partial A_k}\bigg|_{\mathbf{v}_s}
\,.\nonumber
\end{equation}
where
\begin{equation}
P 
= 
M_l\frac{\partial h}{\partial M_l}
+
\rho\frac{\partial h}{\partial \rho}
+
S\frac{\partial h}{\partial S}
+
n\frac{\partial h}{\partial n}
-
h
\,,\nonumber
\end{equation}
as in the Euler relation for pressure.
}

\paragraph{Remark.} One notes many parallels and correspondences among
these equations. Note especially the expected similarities in the equations
for $\mathbf{u}$ and $\mathbf{N}/n$. Recall that 
$\mathbf{A} = -\, \mathbf{N}/n$, 
so that the superfluid velocity is given by $\mathbf{v}_s = \mathbf{u} -
\mathbf{A} = \mathbf{u} + \mathbf{N}/n$.
The evolution of the superfluid velocity is consistently composed as
the sum of these two separate dynamical pieces.

\paragraph{Proof of the Proposition:} The following Hamiltonian $h$ (and
conserved energy) will yield the HVBK equations in the frame
of the normal fluid upon using this Lie-Poisson bracket
\begin{equation}\label{SF-Ham-def}
h = 
\int d^{\,3}x \bigg[ -\frac{1}{2} \rho\, v_n^2 
+
(\mathbf{M} - \rho\mathbf{A})\cdot\mathbf{v}_n
+
\varepsilon_0 (\rho,S,\mathbf{v}_s-\mathbf{v}_n, \,{\omega})
\bigg]\,.
\end{equation}
The variational derivatives of the Hamiltonian $h$ are
computed in this reference frame by using the thermodynamic first law 
(\ref{1stLaw-in-normal-frame}). Namely,
\begin{eqnarray}
\delta{h} 
&=&\int d^{\,3}x
\Big[
\big(\,\mu-\frac{1}{2}v_n^2-\mathbf{A}\cdot\mathbf{v}_n\big)\delta\rho 
+ T\delta{S}
+ \mathbf{v}_n\cdot\delta\mathbf{M}
+ (\mathbf{p} + {\rm curl}\,{\lambda})\cdot\delta\mathbf{u}
\nonumber\\
&&
-\
\big(\mathbf{p} + {\rm curl}\,{\lambda} 
+
\rho\mathbf{v}_n\,\big) \cdot \delta\mathbf{A}
+ 
\big(\mathbf{M} -\mathbf{p} -\rho\mathbf{v}_n 
-  \rho \mathbf{A}\big) \cdot \delta\mathbf{v}_n
\Big].\nonumber
\end{eqnarray}
Here we have used the velocity split 
$\delta\mathbf{v}_s=\delta\mathbf{u}-\delta\mathbf{A}$ and assumed the
boundary condition 
$
{\hat{\mathbf{n}}}\cdot{\omega}
\times\,{\lambda}=0$
when integrating by parts. This boundary condition is satisfied
identically, since  
${\lambda}=\lambda\,{\hat{\omega}}$ in
the HVBK theory. Upon substituting these variational derivatives into the
Lie-Poisson bracket, Corollary \#1 yields the following equations
expressed in the normal fluid reference frame,  
%
\begin{eqnarray}
\partial_t S
&=&
\{S,h\}
=
-\,{\rm div}(S\mathbf{v}_n)
\,,\nonumber\\
\partial_t n
&=&
\{n,h\}
=
-\,{\rm div}(\rho \mathbf{v}_n 
+ \mathbf{p} 
+ {\rm curl}\,{\lambda})
\,,
\nonumber\\
\partial_t\rho
&=&
\{\rho,h\}
=
-\,{\rm div}(\rho \mathbf{v}_n 
+ \mathbf{p} 
+ {\rm curl}\,{\lambda})
\,,\nonumber\\
&&(\,\hbox{Hence, the condition }
n=\rho
\hbox{ is preserved.})
\nonumber\\
\partial_t \mathbf{u}
&=&
\{\mathbf{u},h\}
=
-\,
\nabla(\mu - \frac{1}{2}\,v_n^2 + \mathbf{v}_n\cdot\mathbf{v}_s)
+
\mathbf{v}_n\times {\rm curl}\,\mathbf{u}
\,,\nonumber\\
&&(\,\hbox{Hence, curl}\,
\mathbf{u}=0\
\hbox{is preserved.})
\nonumber\\
\partial_t \mathbf{A}
&=&
\{\mathbf{A},h\}
=
n^{-1}\big(\rho \mathbf{v}_n + \mathbf{p} + 
{\rm curl}\,{\lambda}\big)
\times {\rm curl}\,\mathbf{A}
\,,\nonumber\\
&&(\,\hbox{Hence, }
 \mathbf{v}_\ell
=
 \mathbf{v}_n + \rho^{-1}(\mathbf{p} 
+ 
{\rm curl}\,{\lambda})
\hbox{ when } n=\rho \hbox{ is used.})
\nonumber\\
\partial_t \big(M_j - n A_j \big)
&=&
\{M_j - n A_j,h\}
=
-\,
\partial_k \, T_j^{\,k}
\nonumber
\end{eqnarray}
%
\paragraph{Remarks:} 
\begin{enumerate}
\item [(1.)] Preservation of the condition $n=\rho$ by these
equations allows the introduction of the momentum-carrying field
$\mathbf{A}$ as an independent degree of freedom without introducing
additional material inertia, provided the dynamically preserved condition
$n=\rho$ holds initially. This is reminiscent of the preservation of
Gauss's Law by the continuity equation for mass conservation in a fluid
plasma. 

\item [(2.)]  The curl of the dynamical equation for the field
$\mathbf{A}$ implies the {\bfi vortex line velocity}
\begin{equation}
 \mathbf{v}_\ell
=
-\,\frac{1}{n}\frac{\delta h}{\delta \mathbf{A}}
=
 \bar\mathbf{v} 
+ 
\rho^{-1}{\rm curl}\,{\lambda}
\,,\quad\hbox{where}\quad
\bar\mathbf{v}
=
\mathbf{v}_n + \rho^{-1}\mathbf{p}
=
\rho^{-1}\mathbf{J}
\,.\nonumber
\end{equation}
The velocity $\bar\mathbf{v}$ is the mass averaged velocity. The vortex
slip velocity $\mathbf{s}$ corresponding to the vortex line velocity
$\mathbf{v}_\ell$ is the basis for the phenomenological reactive and mutual
friction forces $\mathbf{f}$ and Rayleigh dissipation function
$R$ in the HVBK system. Namely,
\begin{equation}
\mathbf{s}
=
 \mathbf{v}_\ell -  \mathbf{v}_n
=
\rho^{-1}(\mathbf{p} + {\rm
curl}{\lambda})
\,,\quad\hbox{with}\quad
{\lambda}
=
\lambda\,{\hat{\omega}}
\,.\nonumber
\end{equation}
As expected, this expression agrees with BK [1961] at zero temperature.
Note that {\bfi the renormalized HVBK equations introduce no new parameters}.
\item [(3.)]  The corresponding equation for
$\mathbf{v}_s=\mathbf{u}-\mathbf{A}$ is then obtained as
\begin{equation}
\partial_t \mathbf{v}_s
= 
-\,
\nabla(\mu - \frac{1}{2}\,v_n^2 + \mathbf{v}_n\cdot\mathbf{v}_s)
+
\mathbf{v}_\ell\times {\omega}
\,,
\hbox{ where }
\mathbf{v}_\ell
=
\bar\mathbf{v} 
+ 
\rho^{-1}{\rm curl}\,\lambda\,{\hat{\omega}}
\,.\nonumber
\end{equation}
This may be expressed equivalently in manifestly Galilean invariant form
as
\begin{equation}
\partial_t \mathbf{v}_s
+
(\mathbf{v}_s\cdot\nabla)\mathbf{v}_s
=
-\,
\nabla\big(\,\mu 
- \frac{1}{2}|\mathbf{v}_s - \mathbf{v}_n|^2\big)
+
\mathbf{f}^{\,\prime}
\,,\hbox{ where }
\mathbf{f}^{\,\prime}
=
 (\mathbf{v}_\ell-\mathbf{v}_s)\times {\omega}
\,.\nonumber
\end{equation}
The term $\mathbf{f}^{\,\prime}$ is the {\bfi Hamiltonian contribution to
the reactive vortex force.} This contribution would vanish if the vortex
lines moved with the superfluid velocity.
\item [(4.)]  The stress tensor $T_j^{\,k}=\pi_j^{\,k} + \tau_j^{\,k}$
for total momentum conservation is given by summing
\begin{equation}\label{pi-stress-tensor}
 \pi_j^{\,k} 
=
\big(\,\rho_s\, v_{s\,j}\,v_s^k +\rho_n\, v_{n\,j}\,v_n^k \,\big)
+
P\,\delta_j^{\,k} 
\,,
\qquad
 \tau_j^{\,k} 
= 
\partial_l\,\epsilon_{k\,l\,m}\big(v_{s\,j}\lambda_m\big) 
-
\lambda_j\,\omega^{\,k}
+
{\omega\cdot\lambda}\,\delta_j^{\,k}
\,.\nonumber
\end{equation}
The divergence of $\tau_j^{\,k}$ defines the {\bfi vortex line tension}
$\mathbf{T}$ as
\begin{equation}
\partial_k \tau_j^{\,k}
 = 
-\,
{\omega}\cdot\nabla{\lambda}
+
\nabla({\omega\cdot\lambda})
=
-\,\rho_s\mathbf{T}
+
\nabla({\omega\cdot\lambda})
\,,\nonumber
\end{equation}
In the stress tensor $\pi_j^{\,k}$ the pressure $P$ is defined by the
Euler relation,
\begin{equation}
P = -\,\varepsilon_0 + \mu \rho + TS 
\,,\nonumber
\end{equation}
so that in the normal-fluid frame the pressure satisfies
\begin{equation}
dP = \rho d\mu  + SdT 
-
\mathbf{p}\cdot d(\mathbf{v}_s-\mathbf{v}_n) 
-
{\lambda}\cdot{d}\,{\omega}
\,.\nonumber
\end{equation}
The stress tensor $T_j^{\,k}=\pi_j^{\,k} + \tau_j^{\,k}$ may be derived
by using Corollary \#2 for the Hamiltonian formulation.
\end{enumerate}

\paragraph{Implications of the HVBK vortex dynamics.}

The new Hamiltonian formulation of the renormalized HVBK equations
presented in the Proposition provides a formula for the slip velocity of
a vortex line element in a turbulent superfluid {\sl at finite
temperature}. Namely, for the Hamiltonian $h$ in equation
(\ref{SF-Ham-def}), one finds
\begin{equation}
\mathbf{s}
=\mathbf{v}_\ell
-
\mathbf{v}_n
=
\rho^{-1}\big(\mathbf{p} 
+ 
{\rm curl}\,{\lambda}\big)
\,.\label{slip-vel-HVBK-modified1}
\end{equation}
This formula for $\mathbf{v}_\ell$ recovers the HVBK expression 
in BK [1961] at zero temperature. Otherwise, it provides an 
{\bfi extension to finite temperature} of the HVBK vortex force 
\begin{equation}
\mathbf{f} 
=
(\mathbf{v}_L-\mathbf{v}_s)\times{\omega}
\,,\quad\hbox{with}\quad
\mathbf{v}_L 
=
\mathbf{v}_n
-\,
\rho_s\,( \alpha\,\mathbf{s}
+ 
\beta\,{\hat{\omega}}\times\mathbf{s} )
\,,\label{Phenom-force-HVBK-modified}
\end{equation}
where the {\bfi renormalized vortex slip velocity} is given by
\begin{equation}
\mathbf{s} 
=
\mathbf{v}_\ell-\mathbf{v}_n
=
\frac{\rho_s}{\rho}\,(\mathbf{v}_\ell^0-\mathbf{v}_n)
\,,\quad\hbox{for constant $\rho_s$}
\,.\label{slip-vel-HVBK-modified2}
\end{equation}
The corresponding heating rate $R$ is given by
\begin{equation}
R 
=
(\mathbf{J}-\rho\mathbf{v}_n 
+ 
{\rm curl}\,{\lambda})\cdot{\omega}
\times (\mathbf{v}_L - \mathbf{v}_n)
=
\rho\rho_s\beta\,\omega\,|\mathbf{s}\times{\hat\omega}|^2
\,,\nonumber\label{R-eqn-HVBK-modified}
\end{equation}
which is positive, as it must be.

\section*{References}
\addcontentsline{toc}{section}{References}
\begin{description}

\item
Bekarevich, I. L. and I. M. Khalatnikov [1961]
Phenomenological derivation of the equations of vortex motion in He II,
{\it Sov. Phys. JETP} {\bf 13} 643-646.

\item
Donnelly, R. J. [1999] Cryrogenic fluid dynamics, 
{\it J. Phys.: Condens. Matter} {\bf 11} 7783-7834.

\item 
Gorter, C. J.  and J. H. Mellink [1949]
{\it Physica} {\bf 15} 285.

\item 
Henderson, K. L. and C. F. Barenghi [2000] 
The anomalous motion of superfluid helium in a rotating cavity,
{\it J. Fluid Mech.} {\bf 406} 199-219.

\item Hills, R. N. and P. H. Roberts [1977]
Superfluid mechanics for a high density of vortex lines,
{\it Arch. Rat. Mech. Anal.} {\bf 66} 43-71.

\item 
Holm, D. D.  and B. Kupershmidt [1987]
Superfluid plasmas: multivelocity 
nonlinear hydrodynamics of superfluid solutions 
with charged condensates coupled electromagnetically, 
{\it Phys. Rev. A} {\bf 36} 3947-3956.

\item
Nemirovskii, S. K. and W. Fiszdon [1995]
Chaotic quantized vortices and hydrodynamic processes in superfluid
helium, {\it Rev. Mod. Phys.} {\bf 67} 37-84.

\item
Putterman, S. J. [1974]
{\it Superfluid Hydrodynamics},
North Holland, Amsterdam.

\item 
Reissenegger, A.  [1993] 
The spin up problem in Helium-II
{\it J. Low Temp. Phys.} {\bf 92} 77-106.

\item
Skrbek, L., J. J. Niemela and R. J. Donnelly [1999] 
Turbulent flows at cryogenic temperatures: a new frontier, 
{\it J. Phys.: Condens. Matter} {\bf 11} 7761-7783.

\item 
Sonin, E. B. [1987]
Vortex oscillations and hydrodynamics of rotating superfluids,
Rev. Mod. Phys. {\bf59} 87-155. 1987 

\item 
Volovik, G. E. and V. S. Dotsenko [1979] 
Poisson brackets and continuous dynamics of the vortex lattice in rotating
He-II,
{\it JETP Lett.} {\bf29} 576-579.

\item 
Volovik, G. E. and V. S. Dotsenko [1980] 
Hydrodynamics of defects in condensed media in the concrete cases of
vortices in rotating Helium-II and of disclinations in planar magnetic
substances, 
{\it Sov. Phys. JETP} {\bf58} 65-80.

\end{description}

\end{document}